\documentclass[final,1p,times]{elsarticle} 
\usepackage{graphicx} 
\usepackage{amssymb} 
\usepackage{amsthm} 
\usepackage{lineno} 

\journal{Nuclear Physics A} 
\begin{document} 

\begin{frontmatter} 


\title{System Size and Collision Energy Dependence of $v_2$ 
for Identified Charged Hadrons at RHIC-PHENIX}

\author{Maya Shimomura for the PHENIX Collaboration}

\address{Graduate School of Pure and Applied Sciences, 
Univ. of Tsukuba, Tenno-dai 1-1-1, Tsukuba, Ibaraki, Japan}

\begin{abstract} 
The transverse momentum ($p_{\rm T}$) and centrality dependence of 
the azimuthal anisotropy ($v_{2}$) are measured for charged 
hadrons in different energies and collision species by the 
PHENIX experiment at RHIC. We find that $v_{2}$ divided 
by the participant eccentricity of the initial geometry proportionally 
increases with the number of participants to the 1/3 power 
except at small $N_{\rm part}$ in Cu+Cu at $\sqrt{s_{\rm NN}}$ = 62.4 GeV. 
Taking the eccentricity and quark number scaling into account, there 
is a universal scaling for $v_{2}$ with different energies and collision 
sizes. The results indicate that $v_{2}$ is determined by more than just the 
geometrical eccentricity.  It also depends on the size of collision. 
We also report that both the behaviors of $v_{2}$ and $p_{\rm T}$ 
distributions can be understood from the thermal nature of produced 
particles based on hydro-dynamical behavior.

\end{abstract} 
\end{frontmatter} 


\section{Introduction}
Relativistic heavy ion collisions have been considered as a way 
to create the quark-gluon plasma (QGP), the phase of de-confined quarks
and gluons.  The Relativistic Heavy Ion Collider (RHIC) at Brookhaven
National Laboratory was constructed to 
create and study the QGP. One of the most powerful probes for investigating the 
characteristics of the QGP is measuring the azimuthal anisotropy of the 
charged particles produced by the collisions. The strength 
of the elliptic anisotropy ($v_{2}$) reflects the initial geometrical 
anisotropy, which creates pressure gradients in the collision area.
This pressure gradient transfers a spatial anisotropy into a momentum anisotropy
since particles are pushed harder in the direction of the larger gradients.
Thus, the measured $v_{2}$ reflects the dynamical properties of the dense matter
produced in the collisions.

\section{Motivation}
One of the most remarkable findings at RHIC is that the strength of $v_2$ 
can be described well by hydro-dynamical models in the low transverse 
momentum region ($\sim$ 1 GeV/$c$)~\cite{RP}. In the intermediate 
transverse momentum region (1 $\sim$ 4 GeV/$c$), $v_2$ is consistent with 
quark number ($q_n$) and $KE_{\rm T}$ ( = $m_{\rm T}$ - $m_{0}$) scaling, and the result
supports a quark-recombination model \cite{v2_scaling} .
For a more comprehensive understanding of $v_2$, we have carried out 
systematic measurements of $v_2$, by measuring $v_2$ for identified 
charged hadrons in Au+Au and Cu+Cu collisions at $\sqrt{s_{\rm NN}}$~=~200 and 62.4 GeV, and 
studied the dependence on collision energy, species and centrality. We 
expect that the $v_2$ is determined not only by the initial ellipticity, but is 
also influenced by a finite evolution time which can be related to the 
collision volume.   

\section{Results}
\subsection{Energy, System Size and Species Dependence}
\begin{figure}
\includegraphics[width=\textwidth]{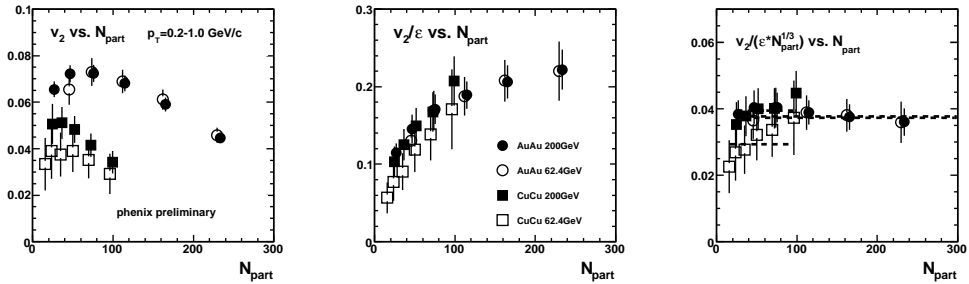}
\vspace{-0.5cm}
\caption{Comparison of integrated $v_{2}$ as a function of $N_{\rm part}$ for four collision systems. 
Left panel shows $v_{2}$ vs. $N_{\rm part}$, middle panel is $v_{2}$/$\varepsilon$ vs. $N_{\rm part}$ and right panel is 
$v_{2}$/($\varepsilon \cdot N_{\rm part}^{1/3}$) vs. $N_{\rm part}$. Closed symbols indicate the results 
of $\sqrt{s_{\rm NN}}$ = 200 GeV, open symbols are for $\sqrt{s_{\rm NN}}$ = 62.4 GeV. Circles indicate Au+Au collisions and 
squares Cu+Cu. The statistical and systematic errors are included in the bars. } 
\label{v2_npart_4systems}
\end{figure}

\begin{figure}[htbp]
\begin{center}
\includegraphics[width=0.65\textwidth]{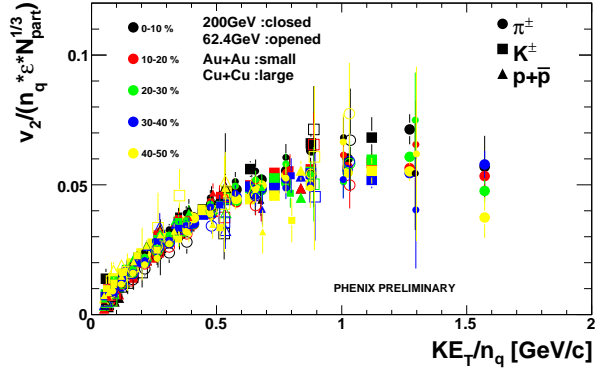}
\vspace{-0.3cm}
\caption{$v_{2}$/($\varepsilon \cdot N_{\rm part}^{1/3} \cdot q_{n}$ ) vs. 
$KE_{\rm T}$/$q_{n}$ for $\pi$/K/p in Au+Au at $\sqrt{s_{\rm NN}}$ = 200 GeV, in Au+Au at $\sqrt{s_{\rm NN}}$ = 62.4 
GeV and in Cu+Cu at $\sqrt{s_{\rm NN}}$ = 200 GeV at five centrality bins for 0-50 \% as 10 \% 
step for each system. There are 45 curves. Applying polynomial fitting, 
$\chi^{2}$/NDF is 2.1. } 
\label{v2_universal_45}
\end{center}
\end{figure}

The left panel in Figure~\ref{v2_npart_4systems} shows $v_2$ vs. $N_{\rm part}$ 
for two collision systems and two collision energies.
The $v_2$ agrees well at $\sqrt{s_{\rm NN}}$ = 200 and 62.4 GeV, but there is a clear 
difference between the Cu+Cu and Au+Au results.
Since Au+Au and Cu+Cu collisions have different initial geometrical 
eccentricities at the same $N_{\rm part}$, applying the eccentricity ($\varepsilon$) scaling
shows that the data follows a universal curve as shown in the middle panel in 
Figure~\ref{v2_npart_4systems}. Here, we use the participant eccentricity,
calculated with a long and short axis determined by the distribution of 
participants at each collision, using a Monte-Carlo simulation based on a Glauber model
and including effects from participant fluctuations. 
The details of the participant eccentricity are described in \cite{par_ecce2} .  
One can see that $v_{2}$/$\varepsilon$ is not a constant and it depends on $N_{\rm part}$ . 
Therefore, $v_{2}$ can be normalized by $\varepsilon$ at the same $N_{\rm part}$, 
but $\varepsilon$ is not enough to determine $v_{2}$ . We newly found that 
$v_{2}$/$\varepsilon$ is proportional to $N_{\rm part}^{1/3}$, and 
as shown in the right panel in Figure~\ref{v2_npart_4systems}  
$v_{2}$/($\varepsilon \cdot N_{\rm part}^{1/3}$) is independent of the collision system
except for small $N_{\rm part}$ in Cu+Cu at $\sqrt{s_{\rm NN}}$ = 62.4 GeV. This exception
indicates this might be a region where the matter has not reached sufficient thermalization,
although the errors are too large to discuss the difference. A scan of collision energies
would be important to further study this effect.   
Figure~\ref{v2_npart_4systems} is for $p_{\rm T}$ = 0.2 - 1.0 GeV/$c$. The 
results for $p_{\rm T}$ = 1.0 - 2.0 and 2.0 - 4.0 GeV/$c$ have the same tendency as well.     
Similar to the result in Au+Au at $\sqrt{s_{\rm NN}}$ = 200 GeV,
$v_{2}$ in Au+Au at $\sqrt{s_{\rm NN}}$ = 62.4 GeV and in Cu+Cu at $\sqrt{s_{\rm NN}}$ = 200 GeV
are mostly consistent with $q_{n}$ + $KE_{\rm T}$ 
scaling for centralities 0 - 50\%. In addition to the fact 
that $v_{2}(p_{\rm T})$ does not depend on collision energy at RHIC energies,
$v_{2}$ normalized by quark number + $KE_{\rm T}$, eccentricity, and 
$N_{\rm part}^{1/3}$ scaling follows a universal curve as shown in Figure~\ref{v2_universal_45}. 
This figure includes the 45 curves for $\pi$/K/p in Au+Au at $\sqrt{s_{\rm NN}}$~=~200 GeV,
in Au+Au at $\sqrt{s_{\rm NN}}$ = 62.4 GeV and in Cu+Cu at $\sqrt{s_{\rm NN}}$~=~200 GeV for the
five centrality bins from 0 - 50\% in 10\% steps.  The $\chi^{2}$/NDF of the
polynomial fitting is 2.1.    
\subsection{Radial Flow Effect with Blast-Wave Fit}
To understand the $N_{\rm part}$ dependence and $KE_{\rm T}$ 
scaling behavior for the $v_2$, we use a Blast-wave model to extract 
dynamical properties of the matter, especially at freeze-out.
This model is a two-parameter model describing a boosted thermal 
source based on relativistic hydrodynamics \cite{BW_phenix}. 
The two parameters, the radial velocity ($\beta_{\rm T}$) and the 
freeze-out temperature ($T_{\rm fo}$), are extracted from the 
invariant cross section data according to the following equation: 

\begin{equation}
{\frac{dN}{m_{\rm T}dm_{\rm T}}} \propto \int_{0}^{R} rdrm_{\rm T}I_{0}
({\frac{p_{\rm T}\sinh{\rho}}{T_{\rm fo}}})K_{1}(\frac{m_{\rm T}\cosh{\rho}}{T_{\rm fo}}) ,
\end{equation}
where $I_{0}$ and $K_{1}$ represent modified Bessel functions with 
$\rho$ the transverse boost $\rho(r)$ = $\tanh^{-1} \beta_{\rm T}(r)$, 
$\beta_{\rm T}(r)$ = $\beta_{\rm s}(\frac{r}{R})$, and $m_{\rm T}$ = $\sqrt{p_{\rm T}^{2} + m^{2}}$.
Re-plotting the measured $p_{\rm T}$ spectra, weighted by measured 
$v_{2}$ ($p_{\rm T}$), we obtain the $p_{\rm T}$ spectra in and out-of plane 
separately for $\pi$/K/p. We use two data sets: Cu+Cu and Au+Au at $\sqrt{s_{\rm NN}}$ = 200 GeV.  
Applying the Blast-wave fitting to the $p_{\rm T}$ spectra in and out-of plane 
separately, the ${\beta}_{\rm T}$ and $T_{\rm fo}$ in and out-of plane are obtained separately. 
From these values, $\beta_{\rm T2}$ = $(\beta_{\rm Tin} - \beta_{\rm Tout})$/$(\beta_{\rm Tin} + \beta_{\rm Tout})$/2 
and $T_{\rm fo2}$ = $(T_{\rm foin} - T_{\rm foout})$/$(T_{\rm foin} + T_{\rm foout})$/2 are calculated.
$\beta_{\rm T2}$ ($T_{\rm fo2}$) indicates the amplitude of the second harmonic of the $\beta_{\rm T}$ ($T_{\rm fo}$) azimuthal distribution. 

\begin{figure}[htbp]
\begin{minipage}[h]{0.48\linewidth}
\begin{center}
\includegraphics*[width=\linewidth]{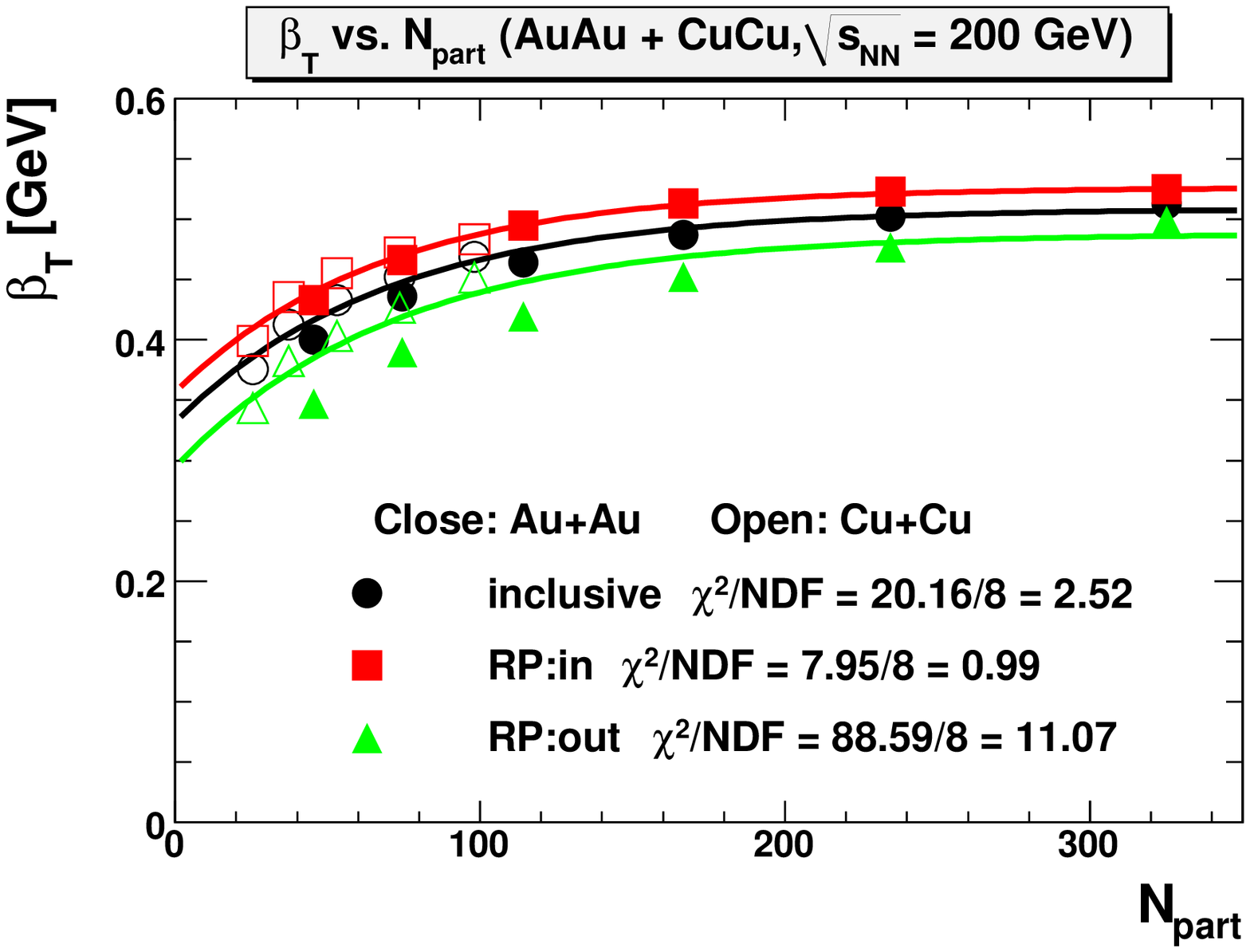}
\vspace{-0.3cm}
\caption{$\beta_{\rm T}$ vs. $N_{\rm part}$ for in and out-of plane in Cu+Cu and Au+Au at 200 GeV. This result is obtained by fitting to PHENIX PRELIMINARY results.}
\label{fig:betaT_vs_Npart_in_out}
\end{center}
\end{minipage}
\hspace{\fill}
\begin{minipage}[h]{0.48\linewidth}
\begin{center}
\includegraphics*[width=\linewidth]{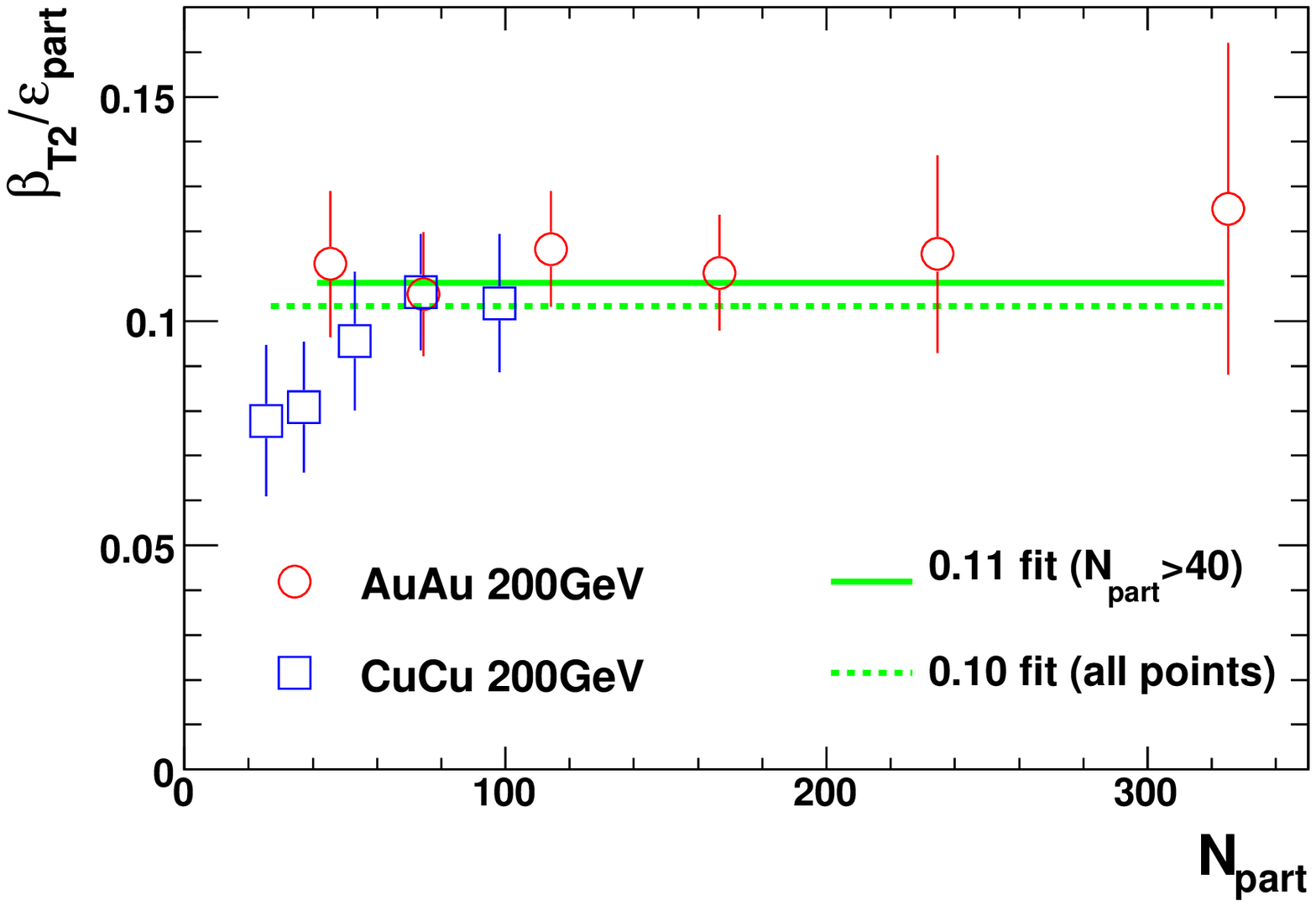}
\vspace{-0.3cm}
\caption{$\beta_{\rm T2}$/$\varepsilon$ vs. $N_{\rm part}$ in Cu+Cu and Au+Au at 200 GeV. This result is obtained by fitting to PHENIX PRELIMINARY results.}
\label{fig:betaT2_ov_epspart_vs_Npart}
\end{center}
\end{minipage}
\end{figure}

\begin{figure}[htbp]
\begin{minipage}[h]{0.48\linewidth}
\begin{center}
\includegraphics[width=\linewidth]{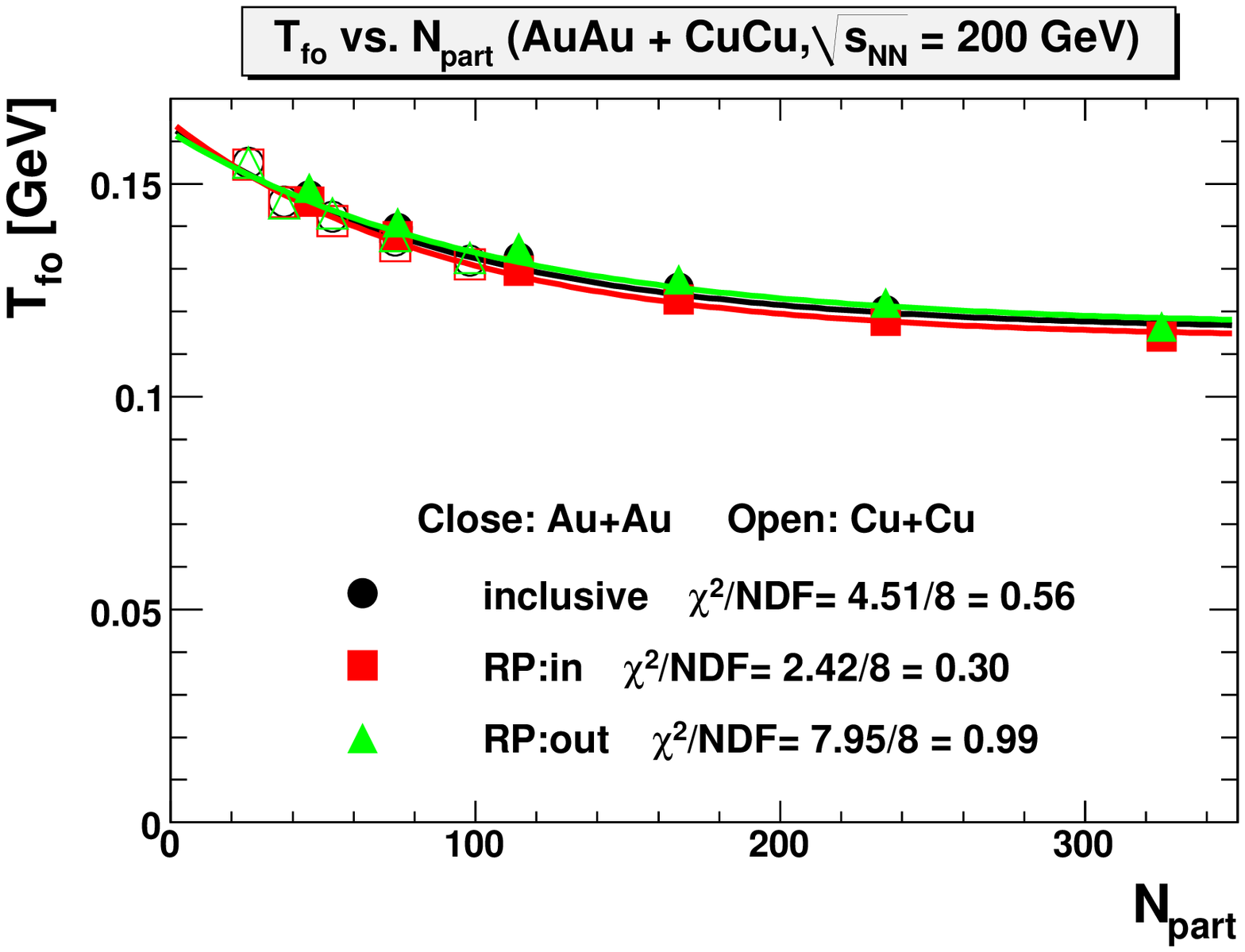}
\vspace{-0.3cm}
\caption{$T_{\rm fo}$ vs. $N_{\rm part}$ for in and out-of plane in Cu+Cu and Au+Au at 200 GeV. This result is obtained by fitting to PHENIX PRELIMINARY results. }
\label{fig:Tfo_vs_Npart_in_out}
\end{center}
\end{minipage}
\hspace{\fill}
\begin{minipage}[h]{0.48\linewidth}
\begin{center}
\includegraphics[width=\linewidth]{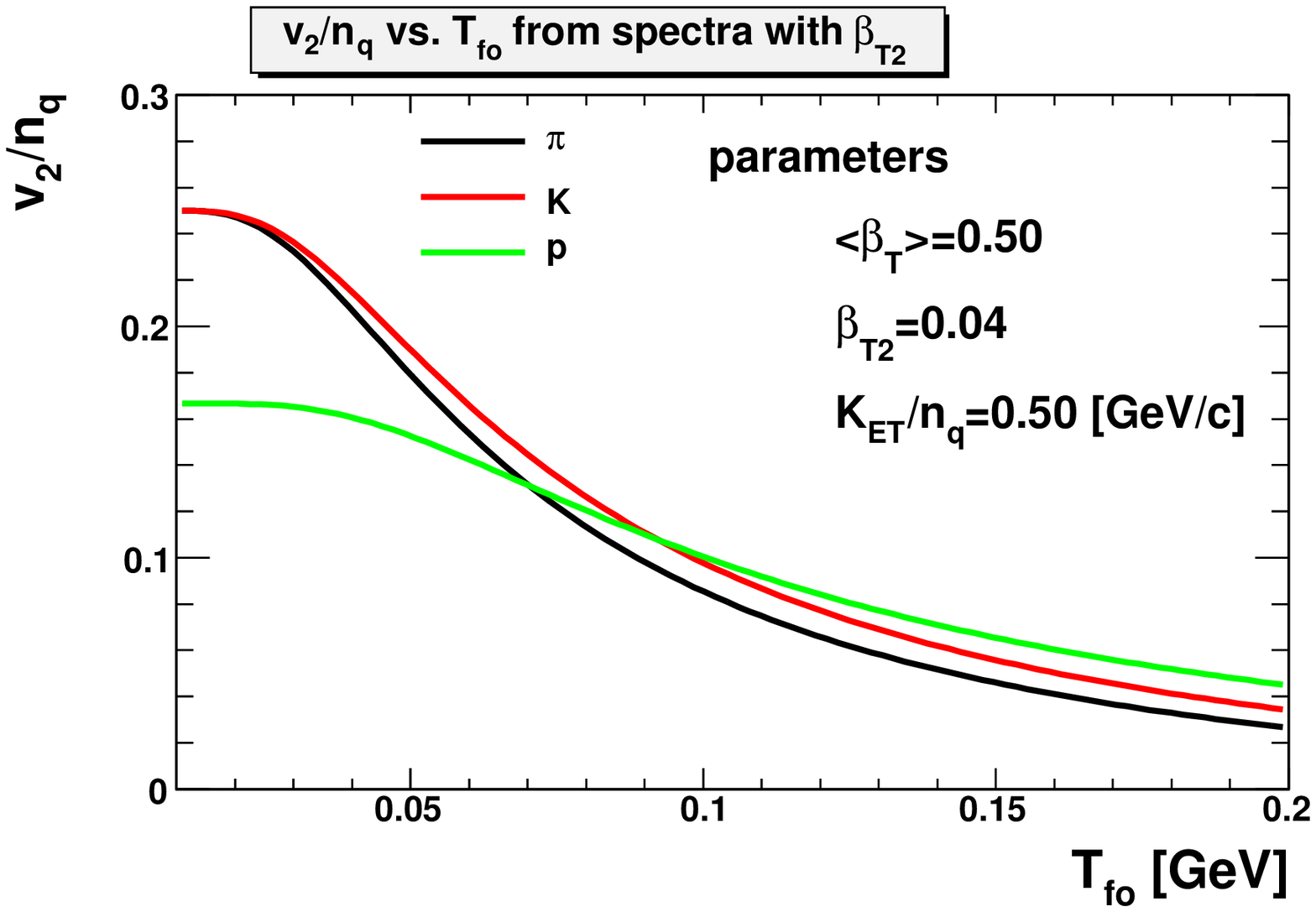}
\vspace{-0.5cm}
\caption{$v_{2}$/$n_{q}$ vs. $T_{\rm fo}$ by the blast-wave calculation with fixed parameters ($\langle \beta_{\rm T} \rangle $ = 0.5, $\beta_{\rm T2}$ = 0.12, $KE_{\rm T}$/$n_{q}$ = 0.5 GeV/$c$). }
\label{fig:v2nq_Tfo_wBW}
\end{center}
\end{minipage}
\end{figure}

Figure~\ref{fig:betaT_vs_Npart_in_out} shows the $\beta_{\rm T}$ vs. 
$N_{\rm part}$ for in and out-of plane in Cu+Cu and Au+Au collisions.
The magnitude of $\beta_{\rm T}$ is clearly different between in and out-of plane, and  
it agrees well between Au+Au and Cu+Cu, especially for the in-plane data. 
The amplitude of the second harmonic, $\beta_{\rm T2}$,  is not the same at the same $N_{\rm part}$ between Cu + Cu and Au + Au. 
When scaled by eccentricity, $\beta_{\rm T2}$/$\varepsilon$ agrees between Au+Au and Cu+Cu and 
it is flat at $N_{\rm part}$ $\geq$ 40 as shown in Figure~\ref{fig:betaT2_ov_epspart_vs_Npart} .
Since $v_2$ is proportional to $\beta_{\rm T2}$ in this model, $v_2$ should be scaled by participant eccentricity. 
However, as shown in Figure~\ref{v2_npart_4systems}, this is not what is seen in the measured $v_2$~results. 
Therefore, this implies that $v_2$ is not determined by only $\beta_{\rm T2}$ 
but includes other effects such as the freeze-out temperature ($T_{\rm fo}$) of radial flow. 
Figure~\ref{fig:Tfo_vs_Npart_in_out} shows the $T_{\rm fo}$ vs. $N_{\rm part}$ for both in and out-of plane.
The magnitude of $T_{\rm fo}$ agrees well between Au+Au and Cu+Cu collision at the same $N_{\rm part}$.
It can be seen that $T_{\rm fo}$ depends on $N_{\rm part}$, and it can influence $v_{2}$ since the larger $T_{\rm fo}$ makes $p_{\rm T}$ spectra flatter.
Figure~\ref{fig:v2nq_Tfo_wBW} shows $v_2$ /$n_{q}$ vs. $T_{\rm fo}$ obtained by the blast wave calculation, fixing parameters to reasonable values ($\langle \beta_{\rm T} \rangle $ = 0.5, $\beta_{\rm T2}$ = 0.04, $KE_{\rm T}$/$n_{q}$ = 0.5 GeV/$c$) and changing only $T_{\rm fo}$. 
In the freeze out temperature region (0.10 $\leq$ $T_{\rm fo}$ $\leq$ 0.16 GeV), it can be seen that $KE_{\rm T}$ scaling is approximately held by this calculation. 
In this figure, going from central to peripheral collisions, namely going from low $T_{\rm fo}$ to high $T_{\rm fo}$, finite deviations among $\pi$/K/p are expected, and $v_2$ for protons becomes larger than that for pions.
This deviation has the same tendency as the $v_2$ data.


\section*{Acknowledgements}
I would like to express my great thanks to the organizers of this conference, Quark Matter 2009, for the opportunity to present these results. I am grateful to my PHENIX collaborators for many helpful discussions.

\end{document}